\documentclass[%
reprint,
 amsmath,amssymb,
 aps,
]{revtex4-2}

\usepackage{graphicx}
\usepackage{dcolumn}
\usepackage{bm}
\usepackage{xcolor}
\begin{document}

\preprint{APS/123-QED}

\title{Associative ionization in a dilute ultracold $^7$Li gas probed with a hybrid trap}

\author{N. Joshi,$^{1,\dagger,*}$ Vaibhav Mahendrakar,$^{1,\dagger}$ M. Niranjan,$^{1,2,\dagger}$ Raghuveer Singh Yadav,$^1$\\ E Krishnakumar,$^1$ A. Pandey$^3$,  R Vexiau$^3$, O. Dulieu$^3$, S. A. Rangwala$^{1,}$}

 \email{Email: njoshi@rri.res.in}
 \email{sarangwala@rri.res.in\\$\dagger$ Contributed equally}
 
\affiliation{%
 $^1$Raman Research Institute, C. V. Raman Avenue, Sadashivanagar, Bangalore 560080, India\\
 $^2$Laboratoire Charles Fabry, Institut d’Optique, CNRS, Université Paris-Saclay, 91127 Palaiseau,
France\\
 $^3$Laboratoire Aimé Cotton, CNRS, Université Paris-Saclay, Orsay, 91400, France}%

\date{\today}

\begin{abstract}
The formation of Li$_2^+$ and subsequently Li$^+$ ions, during the excitation of $^7$Li atoms to the $3S_{1/2}$ state in a $^7$Li magneto optical trap (MOT), is probed in an ion-atom hybrid trap. Associative ionization occurs during the collision of Li($2P_{3/2}$) and Li($3S_{1/2}$) ultracold atoms, creating Li$_2^+$ ions. Photodissociation of Li$_2^+$ by the MOT lasers is an active channel for the conversion of Li$_2^+$ to Li$^+$. A fraction of the Li$_2^+$ ions is long lived even in the presence of MOT light. Additionally,  rapid formation of Li$^+$ from Li$_2^+$ in the absence of MOT light is observed. Resonant excitation of ultracold atoms, resulting in intricate molecular dynamics, reveals important processes in ultracold dilute gases. 

\end{abstract}

\maketitle

\begin{figure}[h]

\centering
\hspace{-0.3 cm}
\includegraphics[scale=0.105]{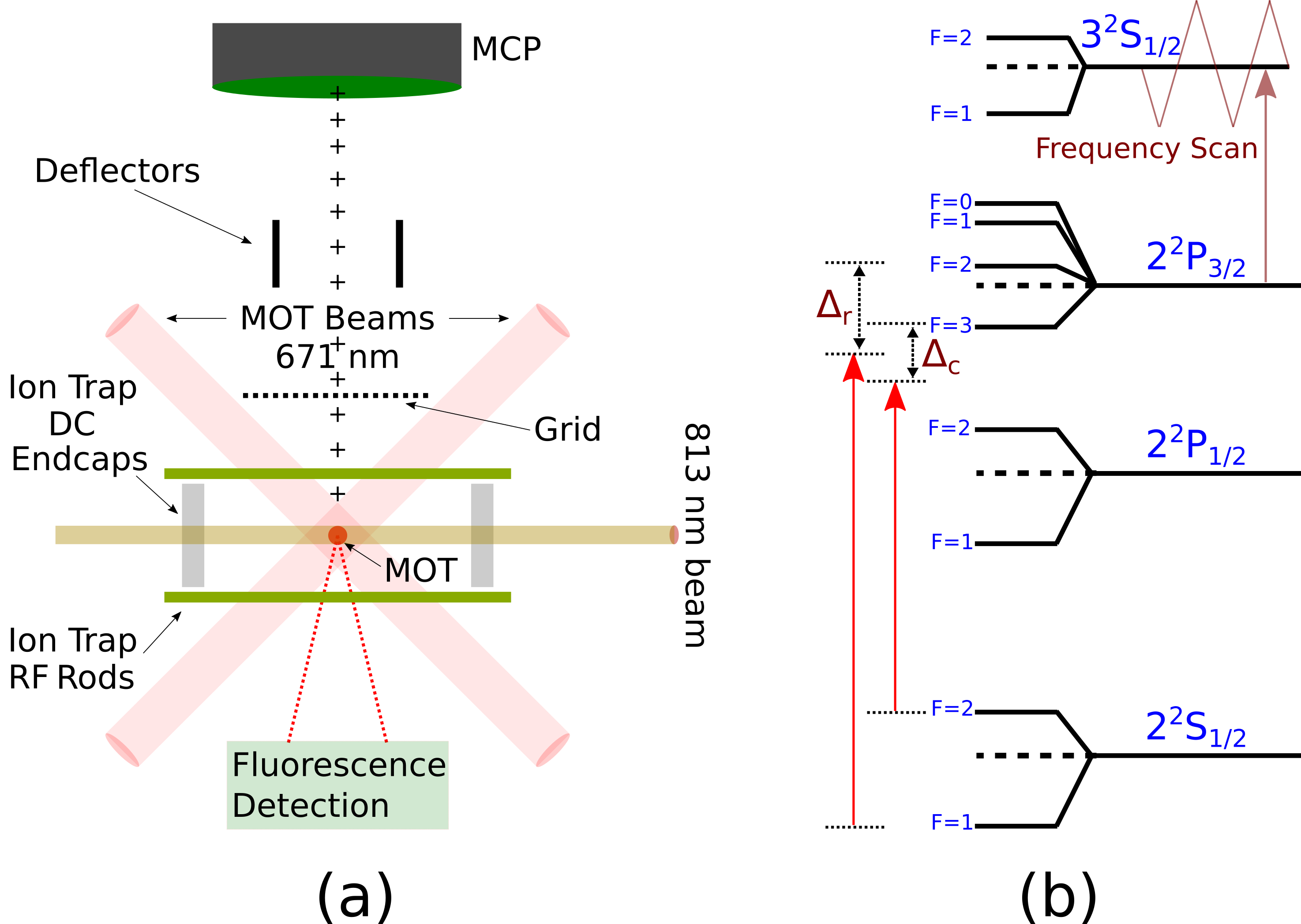}

\caption{(a) Schematic representation of the experimental setup. The MOT is positioned at the ion trap center. (b) Energy level diagram of $^7$Li for the MOT operation and for the excitation of the MOT atoms to the $3S_{1/2}$ state. The $3S_{1/2} \rightarrow 2P_{3/2}$ fluorescence is monitored by scanning the 813~nm laser across the $2P_{3/2}  \rightarrow 3S_{1/2}$ transition frequency (see SM for details). See SM for details of the fluorescence detection optical setup.\label{fig:ExperimentDiagram}}\end{figure}

\begin{figure}[h]

\centering

\includegraphics[scale=0.6]{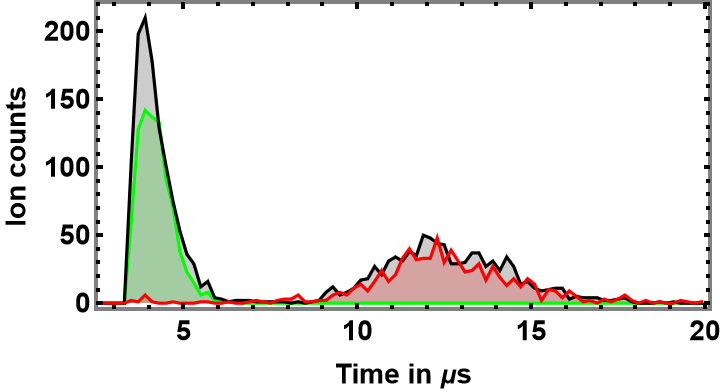}

\caption{ToF spectrum, recorded in configuration A (see SM), of the ions created by the 813~nm laser (black curve) and the UV LED (green curve). The red curve shows a loss of the 813~nm generated ions, which appear at the same ToF as the UV LED generated ions, in the presence of a parametric frequency $f_p$(Li$^+$)=149~kHz (Table \ref{tab:fpe}).
\label{fig:813nmParametric}}
\end{figure}

\hspace{-0.35 cm}Experiments with hybrid ion-atom traps, which combine laser-cooled ions with ultracold neutral atoms have opened up new possibilities for studying complex interactions and processes \cite{ smith2005cold,grier2009observation,zipkes2010trapped,hall2011light,ravi2012combined, PhysRevA.87.052715,jyothi2015hybrid,lopez2015sympathetic,hudson2016sympathetic,meir2018experimental,jyothi2019hybrid,hirzler2020experimental,schmidt2020optical,veit2021pulsed, RevModPhys.91.035001}. In such systems the products of ion-atom collisions can be trapped, leading to unprecedented details of interaction outcomes \cite{ravi2012cooling,dutta2018cooling,jyothi2016photodissociation,sikorsky2018spin, stoecklin2016explanation,mahdian2021direct,pinkas2023trap,weckesser2021observation,zuber2022observation}. One can measure collision rates and branching ratios of competing channels \cite{grier2009observation,lee2013measurement,dutta2020measurement, haze2018cooling,joger2017observation,kwolek2019measurement,benshlomi2020,ben2021high,feldker2020buffer,krukow2016energy,mohammadi2021life,xing2022} leading to the discovery of processes previously undetected in beam experiments.\\
In this article we use a hybrid trap as a tool to probe the formation of Li$_{2}^{+}$ and subsequently Li$^{+}$ ions when a low intensity and narrow linewidth laser at 813~nm, resonant with the $2P_{3/2} \rightarrow 3S_{1/2}$ transition in $^{7}$Li is incident on a magneto optical trap (MOT) of $^{7}$Li. These processes have not been reported in former ultracold $^7$Li atom experiments, and in beam experiments that employed the $2P \rightarrow 3S$ excitation \cite{iu1995instrumentation,stevens1995hyperfine,rubbmark1981dynamical,littman1978field}, as an intermediate step, for producing lithium Rydberg atoms. Below, we report a series of measurements, and employ electronic structure calculations on Li$_2$ and Li$_2^+$ as well as computation of Li$_2^+$ photodissociation rates for understanding our experimental observations.\\
Our hybrid trap experiment (Fig. \ref{fig:ExperimentDiagram}(a)), consists of a $^{7}$Li MOT created at the center of a linear Paul trap with 4 linear, and 2 dc endcap electrodes. The MOT typically contains 1.7$\times$10$^{6}$ atoms, with a density of 5.6$\times$10$^{14}$~atoms/m$^3$ at a temperature of 500~$\mu$K. The values of the red detuning of the cooling laser ($\Delta_{c}$) from the $2S_{1/2}(F=2) \rightarrow 2P_{3/2}(F=3)$ transition, and of the repumping laser ($\Delta_{r}$) from the $2S_{1/2}(F=1) \rightarrow 2P_{3/2}(F=2)$ transition (Fig \ref{fig:ExperimentDiagram}(b)), are $\Delta_{c}\approx -38$~MHz, and $\Delta_{r} \approx -22$~ MHz. The total cooling and repumping intensities are $\approx 77$~mW/cm$^{2}$ and $\approx 25$~mW/cm$^{2}$ respectively and the magnetic field gradient is 13.6 Gauss/cm. More details on the MOT operation are provided in the Supplemental material (SM). 
The Paul trap is operated with an angular frequency and amplitude of rf ($\Omega_{rf}$, $V_{rf}$) applied to one diagonal pair of electrodes while the other pair is grounded \cite{drakoudis2006instabilities,sinhal2023molecular}. The end cap electrodes are at constant potential $V_{ec}$. The ions, once created are trapped in the ion trap and extracted onto a microchannel plate (MCP) by switching off the rf voltage at zero phase, while simultaneously applying a brief, negative high voltage (HV) pulse to the grid electrode (see Fig. \ref{fig:ExperimentDiagram} (a)). Ions hitting the MCP build a time of flight (ToF) mass spectrum and can be counted (Fig. \ref{fig:813nmParametric}).\\
The experiment is conducted as follows. The $^7$Li MOT is loaded under standard conditions while the ion trap is kept on with operating parameters $V_{rf}=70$~V, $\Omega_{rf}/2\pi=1$~MHz, $V_{ec}=4$~V (Configuration A, Table \ref{tab:table1} in SM). When an 813~nm narrow linewidth laser driving the $2P_{3/2}\rightarrow 3S_{1/2}$ ($F=1$ or $F=2$) transition of    

\begin{figure}[h]
\centering
\includegraphics[scale=0.45]{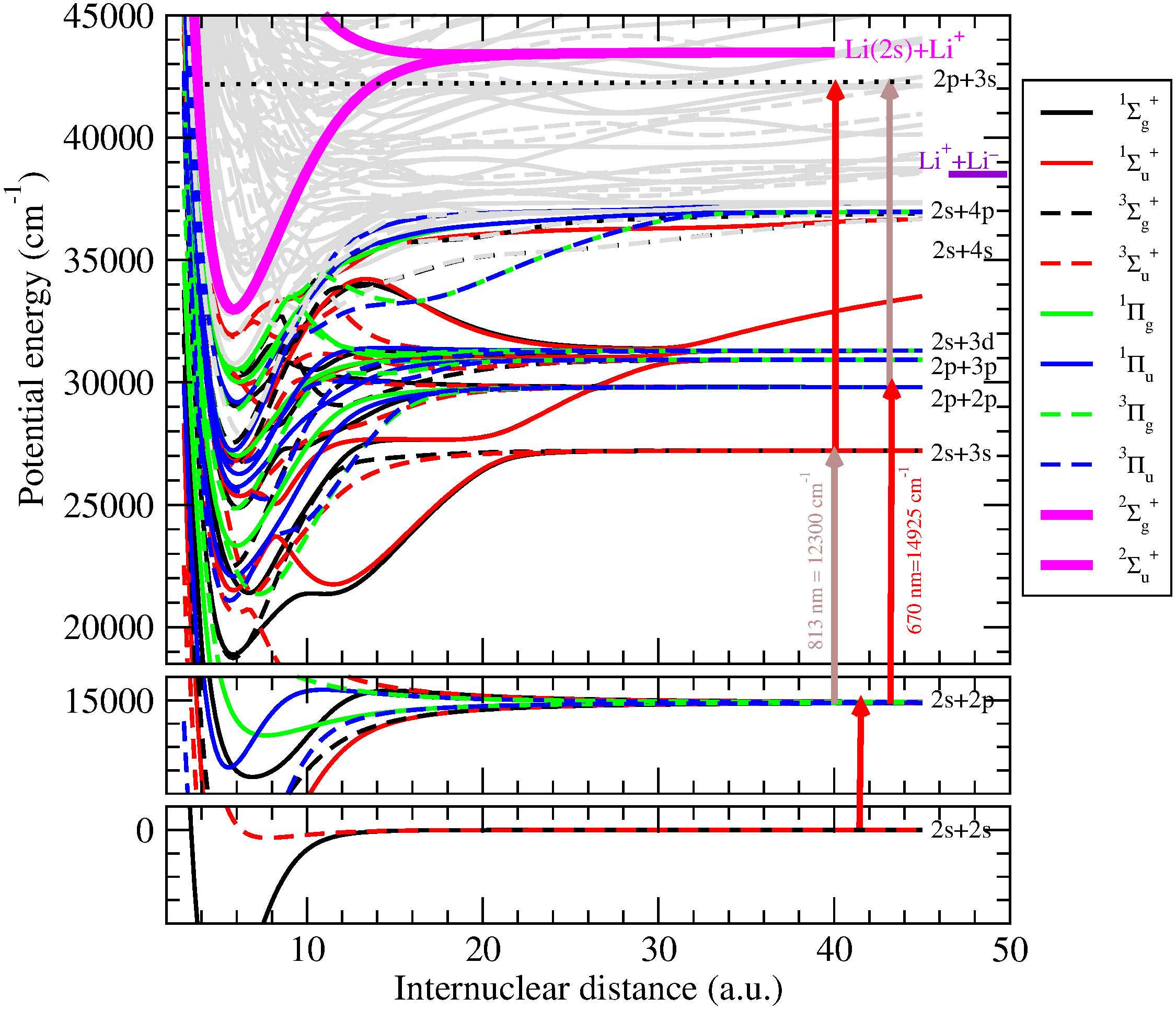}
\caption{PECs of Li$_2$ ground and excited electronic states up to the energy range of the lowest Li$_2^+$ PECs (thick full magenta lines), computed according to \cite{aymar2005,vexiau2017}. The energy of the Li$^+$+Li$^{-}$ is indicated for completeness. The grey curves illustrate the complexity of the PECs of highly excited states, which are irrelevant at short distances when they enter the ionization continuum. The arrows picture the possible excitation pathways of the Li atoms. 
\label{fig:NeutralPEC}}
\end{figure}

\hspace{-0.35 cm}$^7$Li MOT atoms is turned on (see Fig. 1(b) and SM), and the above extraction sequence is followed, a mass spectrum exhibiting two peaks is observed (Fig. \ref{fig:813nmParametric} black curve). When only the MOT is operational, no ions are produced and detected upon extraction. The narrow peak at $\approx 4 \mu$s is assigned to the Li$^+$ signal, as it coincides with the peak (green curve in Fig. \ref{fig:813nmParametric}) recorded when a sole UV LED (M340L4 Thorlabs) is employed to directly photoionize the Li($2P_{3/2}$) atoms in the MOT. The broad distribution around 12~$\mu$s could be induced by any ion with a mass larger than that of the Li$^+$ monomer, including Li$_m^+$ ions with $m>1$. Due to the pulsed grid extraction, the ToF of the detected ions does not scale as the square root of ion masses. We, therefore, use the parametric excitation (PE) technique \cite{major2005charged,zhao2002parametric} to characterize the broad distribution. A small perturbation with a frequency $f_p$ is applied to the trapped ions. The energy transfer to the ion is maximum when $f_p$ is twice the secular frequency $f_s$ of the trapped ions. A PE resonance is detected when the ions of a particular charge-to-mass ratio $(Q/M)$ are rapidly ejected from the trap, leaving ions with different $(Q/M)^{\prime}$ unaffected. By tuning $f_p$, the trapped ions for various $(Q/M)$ can be selectively emptied (see SM). We first load the trap from MOT atoms by pulsing the UV LED for a short time interval and determine the PE resonance, $f_p$(Li$^+$), for Li$^+$ by tuning $f_p$ in the configuration A (see Table \ref{tab:table1} in SM). To determine PE resonance frequency for Li$^+_{m}$ ions created when applying the 813~nm laser, we eliminate Li$^+$ by tuning the extraction pulse on the grid electrode such that the trapped Li$^+$ ions do not reach the detector while Li$^+_{m}$ ions do (configuration B, Table \ref{tab:table1} in SM). Then $f_p$ is tuned to locate the PE resonance, $f_p$(Li$^+_m$), for Li$_m^+$. The ratio $f_p$(Li$^+$)/$f_p$(Li$_m^+$) is determined for different values of $V_{ec}$ and converges to 2 when $V_{ec}\rightarrow 0$  (see Table \ref{tab:fpe} and \cite{collings2003resonant}). This establishes that the broad distribution peaked at 12~$\mu$s in Fig. \ref{fig:813nmParametric} contains Li$^+_2$ ions.\\
We investigate the formation mechanism of the Li$^+$ and Li$^+_{2}$ ions which appears to be a complex process induced by the excited Li atoms. The ionization energy of the $2P_{3/2}$ and $3S_{1/2}$ levels does not allow the creation of Li$^+$ by single photon ionization of lithium atoms from these levels by the 670~nm or 813~nm lasers. In addition, multiphoton ionization is unlikely as there is no resonant intermediate state to enhance ionization \cite{vsibalic2017arc,oxley2010frequency}. This raises the issue of the origin of the Li$^+$ ions. First, in configuration A (see SM), we observe that in the presence of a parametric drive at 149~kHz (Table \ref{tab:fpe}) there is a loss of the 813~nm generated ions that have the same ToF as the UV LED generated Li$^{+}$ ions (red curve in Fig. \ref{fig:813nmParametric}), reconfirming that this fraction of the ions is indeed Li$^{+}$. We further test whether the 813~nm
excitation generates only Li$^+_{2}$ to begin with. In configuration A while shining the 813~nm laser, the application of $f_p$(Li$_m^+$)=65~kHz eliminates both the Li$^+$ and Li$_2^+$ peaks in the ToF spectrum of Fig. \ref{fig:813nmParametric}, except for rare counts at the ToF of Li$^+$. In contrast, the Li$^+$ ions created with the UV LED are unaffected by the 65~kHz excitation, as determined in a separate experiment. This strongly indicates that the Li$^+$ ions are created from Li$_2^+$ and the ions in the broad peak are Li$_2^+$ exclusively. For this experiment, the PE of the Li$_2^+$ ions has to be rapid so that the Li$_2^+$ ions are ejected from the trap before a sizable fraction of Li$^+$ forms. This required the PE amplitude of Li$_2^+$ to be increased by $\approx$ 9.3 times the typical value (Table \ref{tab:fpe}).\\
To identify the process leading to Li$_2^+$ production, we calculate the potential energy curves (PECs) for Li$_2$ and Li$_2^+$ (Fig. \ref{fig:NeutralPEC}), using the same methodology as reported earlier \cite{aymar2005,vexiau2017}. A colliding pair of $2P_{3/2}$ and $3S_{1/2}$ atoms have enough energy to reach the Li$_2$ ionization threshold undergoing associative ionization (AI) \cite{dulieu1991theoretical,dulieu1994doubly,dulieu1994application,polak1981observation,polak1980observation,urbain1991dynamics,hellfeld1978observation,babenko1995associative,gabbanini1991associative,mcgeoch1988associative,ono1970mass}, which occurs due to autoionization of the two interacting excited atoms scattering along several PECs which asymptotically correlate to the $2P_{3/2}+3S_{1/2}$ dissociation limit. A fraction of these colliding pairs generates Li$_2^+$, and the rest radiatively decays back to ground atomic states. By tuning the 813~nm laser over a range of frequencies $> 100$~GHz across the $2P_{3/2} \rightarrow 3S_{1/2}$ resonance, the Li$_2^+$ signal manifests only when the sum of the photon energies of the 670~nm and 813 nm laser equals the energy difference between the $2S_{1/2}$ and $3S_{1/2}$ atomic levels. This weighs against other processes that involve, as an intermediate step, the photoassociation (PA) of neutral Li$_2$ in a 

\begin{table}[h]
\caption{Comparison of the most prominent resonant frequency for Li$^+$ and Li$_{m}^{+}$ measured by varying the frequency of the parametric drive ($f_p$), in steps of 1 kHz and recording the number of ions reaching the MCP at each frequency value. The peak-to-peak parametric drive amplitude, in units of milli volts ($mV_{pp}$), is kept to be 180 $mV_{pp}$ and 67.5 $mV_{pp}$ respectively, for the Li$^{+}$ and the  Li$_{m}^{+}$.}
\begin{ruledtabular}
\begin{tabular}{cccc}
V$_{ec}$&
$f_p$(Li$^+$)&
$f_p$(Li$_m^+$) &
$f_p$(Li$^+$)/$f_p$(Li$_m^+$)\\
(V)&
(kHz)&
(kHz) &
 \\
\colrule
4 & 149 $\pm$ 0.5 & 65 $\pm$ 0.5 & 2.29 $\pm$ 0.03\\
2 & 155 $\pm$ 0.5 & 72 $\pm$ 0.5 & 2.15 $\pm$ 0.02\\
1 & 158 $\pm$ 0.5 & 76 $\pm$ 0.5 & 2.08 $\pm$ 0.02\\
\end{tabular}
\end{ruledtabular}
\label{tab:fpe}
\end{table}

\begin{figure}[h]
\centering
\includegraphics[scale=0.36]{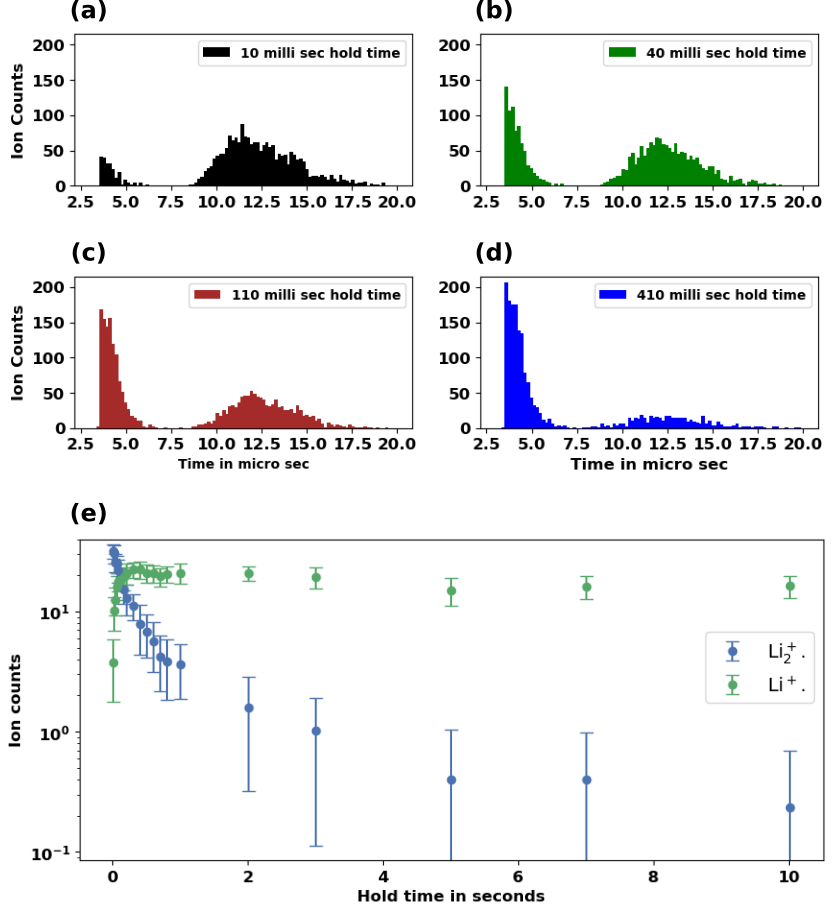}
\caption{(a-d) ToF spectra of the ions generated by the 813~nm laser pulse, for increasing holding times in the presence of the MOT. Hold time is measured with respect to the rising edge of the 813 nm pulse. In the ToF spectra, we identify the signal below and above 7~$\mu$s as Li$^+$ and Li$^+_2$ respectively (see SM). For each holding time, 60 measurements are performed and used to construct the histograms. (e) Variation of the mean number of Li$^+$ and Li$_2^+$ ions per shot with the holding time. The error bars represent one standard deviation.\label{fig:TimeEvolve}}
\end{figure}

\begin{figure}[h]
\centering
\includegraphics[scale=0.21]{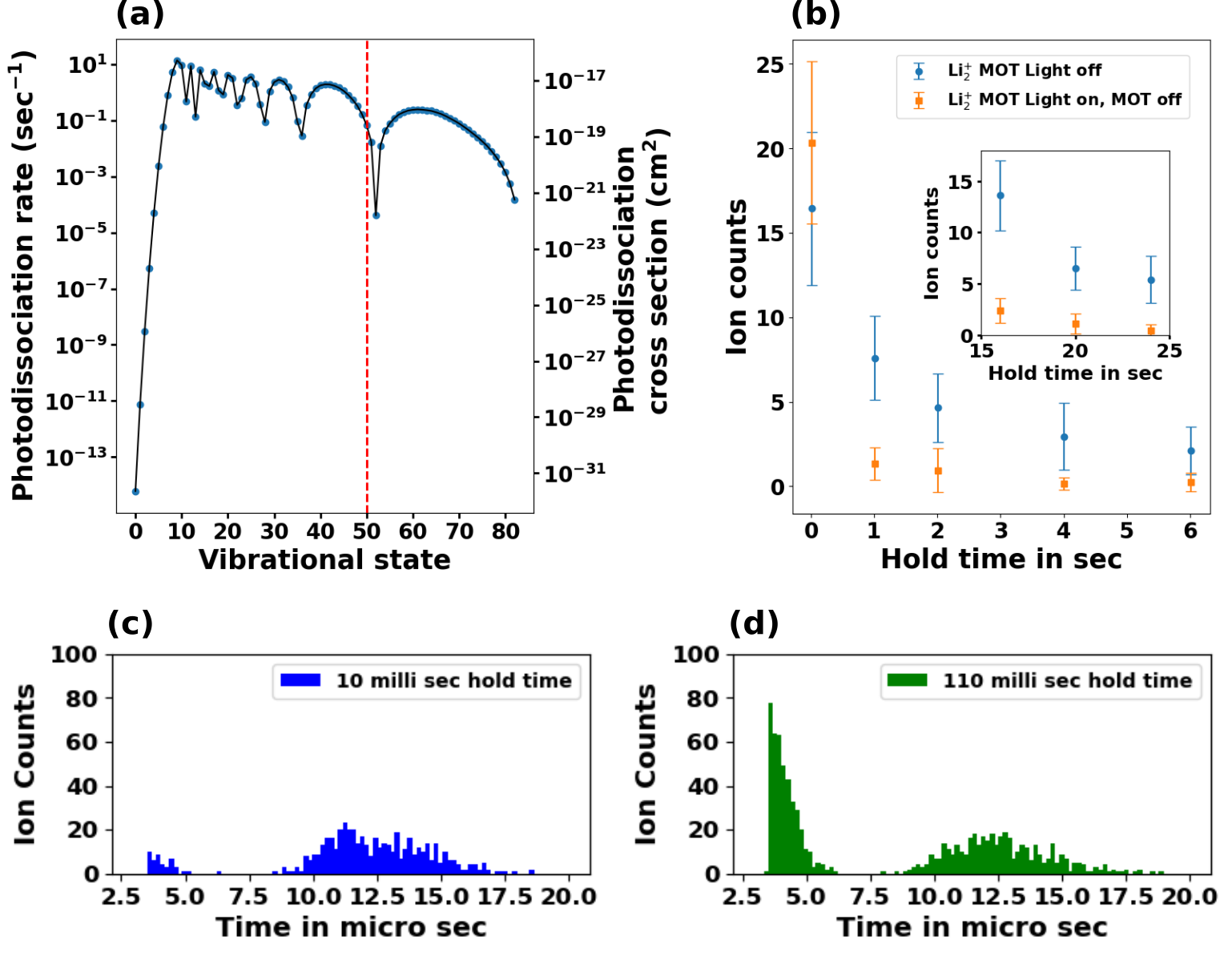}
\caption{(a) Calculated Li$_2^+$ PD cross section and non-thermalized rate at 670~nm for the vibrational levels of the $^{2}\Sigma_{g}^{+}$ ground state toward the dissociation continuum of the lowest $^{2}\Sigma_{u}^{+}$ state. The red dashed line marks the $v_g=50$ vibrational level, which is the highest level that can be populated when entering by the $2P_{3/2}+3S_{1/2}$ scattering channel. (b) Lifetime measurements of Li$_{2}^{+}$ made in configuration C (see Table \ref{tab:table1} in SM) in the presence and absence of the MOT laser. The inset shows the existence of long lived Li$_2^+$ ions in the presence of MOT light. Panels (c) and (d) show the change in ToF spectra indicating conversion of Li$_2^+$ to Li$^+$ even in the absence of MOT light. The MOT level was not ensured to be the same in (c) and (d) as we just wanted to depict change in the ToF.\label{fig:MolLifetime}}
\end{figure}

\hspace{-0.35 cm}specific bound rovibrational level which can either be photoionized (PI) directly (PA-PI \cite{blange1997vibrational}) at short range or can be further excited to an autoionizing state, commonly referred to as photoassociative ionization (PAI) \cite{weiner2003cold}. From Fig. \ref{fig:NeutralPEC}, it is impossible to identify the molecular states responsible for AI, as the relevant energy zone close to the $2P_{3/2}+3S_{1/2}$ dissociation limit corresponds to highly-excited states with adiabatic PECs which have a poor accuracy at short distances, as they cross the ionization threshold (grey curves in Fig. \ref{fig:NeutralPEC}). Based on the calculated PECs, we infer that the Li$_2^+$ ions are created in the $X^2\Sigma_g^+$ ground electronic state with a vibrational level not higher than $v_g=50$, with a binding energy of about 1000~cm$^{-1}$ or more. We measured the total number of created Li$_2^+$ and Li$^+$ ions for constant MOT density, which is linear with the intensity of the 813~nm laser, indicating that a single photon is involved (see SM and \cite{trachy2007photoassociation}). For a constant 813~nm laser intensity, the number of detected ions varies quadratically with the MOT density, which is the signature of a two-atom process (see SM and \cite{gould1988observation}).\\
As invoked above, the Li$^+$ yield must originate from the Li$_2^+$ ions. Two options are plausible: (i) photodissociation of Li$_{2}^{+}$ by the MOT lasers \cite{jyothi2016photodissociation} or by the 813~nm laser; (ii) charge exchange with neutral Li atoms \cite{pandey2024ultracoldchargedatomdimercollisions}. Another possibility of creating Li$^+$ which does not involve the formation of Li$_2^+$ in the $X^{2}\Sigma_{g}^{+}$ state, is the autoionization of colliding $3S_{1/2}+3S_{1/2}$ atoms along the lowest $A^{2}\Sigma_{u}^{+}$ dissociative Li$_2^+$ PEC. This was the major pathway for the formation of atomic ions in a previous study employing a similar scheme for exciting Rubidium MOT atoms and was referred to as photoassociative dissociative ionization (PADI) \cite{trachy2007photoassociation}.\\
We study the ion population dynamics with the ion trap and HV extraction pulse set up in configuration A (see SM), which is suitable for trapping and detecting both Li$^+$ and Li$_2^+$. The MOT atoms are irradiated by a 10~ms pulse of the 813~nm laser with 5~mW/cm$^2$ intensity. The created ions are held in the ion trap, in the presence of MOT atoms, for varying holding times of duration 10~ms and longer, and extracted toward the MCP (Fig. \ref{fig:TimeEvolve}). In panels (a)-(d), we see that the population of Li$^+$ increases with increasing holding times, while the Li$_2^+$ population decreases. Panel (e) shows that after a fast decrease (increase) of the Li$_2^+$ (Li$^+$) signal, the mean number of each ion species stabilizes over long holding times (see SM for the counting methodology). This experiment confirms that as time progresses, the Li$_2^+$ ions are converted into Li$^+$ ions. The measured small number of Li$^+$ for the 10~ms holding time measurement (Fig. \ref{fig:TimeEvolve}a) reflects the negligible population of the $3S_{1/2}+3S_{1/2}$ scattering channel, which would directly produce Li$^+$ via PADI.\\
We next investigate the photodissociation (PD) of Li$_2^+$ by the MOT lasers. In Fig. \ref{fig:MolLifetime}(a) we display the computed PD cross section and rate \cite{miyake2011rovibrationally,jyothi2016photodissociation} at 670~nm of the Li$_2^+$ ground electronic state (X$^{2}\Sigma_{g}^{+}$) vibrational levels 

\hspace{-0.35 cm}toward the dissociation continuum of the  $A^{2}\Sigma_{u}^{+}$ state, using the PECs of Fig. \ref{fig:NeutralPEC}. We see that the few lower vibrational levels cannot be photodissociated by the 670~nm light due to a large binding energy. If populated, these ions will have a long lifetime in the trap, unlike the Rb$_2^+$ case where all levels were affected by the MOT light \cite{jyothi2016photodissociation}.\\
To probe the  photodissociation of Li$_2^+$, we perform measurements similar to those of Fig. \ref{fig:TimeEvolve}, operating the ion trap in configuration C (Table \ref{tab:table1} in SM), which traps only Li$_2^+$. The MOT is loaded for 20~s, and the 813~nm laser with 5 mW$/$cm$^2$ intensity is switched on for 10~ms just before switching off the repumper and emptying the MOT. In 10~ms, more than 70$\%$ of the MOT atoms leave the volume of the overlapping MOT beams as determined experimentally using the release and recapture method \cite{chu1985three}. The ions created by the 813~nm laser are held in the ion trap for various duration either in the presence or in the absence of the MOT cooling light, and are extracted toward the MCP. The results clearly indicate that the molecular ions have a much shorter lifetime in the presence of MOT light, which then contributes to the loss of Li$_2^+$ (Fig. \ref{fig:MolLifetime}(b)). We neglect the PD of Li$_2^+$ due to the 813~nm laser as its intensity is very small compared to the MOT lasers and also it is pulsed only for a short time. Further, a laser with a wavelength $\leq$ 444~nm can photodissociate even the lowest vibrational level of Li$_2^+$ which, due to larger binding energy, is immune from PD by the 670~nm light. We indeed observe a lower lifetime if the Li$_2^+$ ions are exposed to both the 670~nm and a 420~nm laser.\\
We further investigated the presence of long lived molecular ions in the presence of the MOT light. The pulse width of the 813~nm laser was increased to 500~ms to produce a large number of ions so that, statistically, a sizable fraction of the ions may be generated in low vibrational levels and thus with a low photodissociation rate. The results of the measurements are shown in the inset of Fig. \ref{fig:MolLifetime}(b). We present data only at large holding times because the counting of ions at small holding times is not possible due to MCP saturation.\\
From Fig. \ref{fig:MolLifetime}(b), we also observe a rapid decrease in the number of Li$_2^+$ in going from 10~ms to 1~s holding time even in the absence of MOT light. Since this loss is not due to the trap lifetime, we probe the short timescale dynamics in the absence of MOT light. The measurements, similar to the ones in Fig. \ref{fig:MolLifetime}(b), were made in configuration A which traps both Li$^+$ and Li$_2^+$. We present two measurements, shown in Fig. \ref{fig:MolLifetime}(c) and \ref{fig:MolLifetime}(d): they indicate a change in the ToF histograms at different hold times and signals the presence of another process, which leads to the initial, non-optical rapid conversion of Li$_2^+$ to Li$^+$. The cause of this rapid conversion is yet to be explored as it cannot be further probed in the present experiment. We theoretically evaluated the rates of charge exchange of Li$_2^+$ with background Li atoms by considering the Langevin collision model \cite{mcdaniel1964collision,harter2014cold}. However, the collision rates are small with both the background as well as MOT atoms to explain the fast conversion of Li$_2^+$ to Li$^+$ in Fig. \ref{fig:MolLifetime}(b) and Fig. \ref{fig:TimeEvolve}(e). We have also observed the formation of Li$_2^+$  ions in driving the $2P\rightarrow 3S$ transition in a $^6$Li MOT by the 813~nm laser. Signatures of the formation of Li$_2^+$ were reported earlier in a $^6$Li MOT illuminated with a femtosecond laser having a 750~nm-820~nm spectral range \cite{kurz2021kinematically}.\\
\hspace{-0.35 cm}In summary, we have shown that the simple process of resonant excitation from the first excited state of Li atoms in a dilute ultracold gas leads to rich dynamics with a large probability of formation of molecular and atomic ions. This has important implications for cold dilute and degenerate gas experiments, where photo-excitation is used \cite{ewald2019observation,dieterle2020inelastic,deiss2021long}, and for Rydberg atom experiments in general \cite{iu1995instrumentation,stevens1995hyperfine,rubbmark1981dynamical,littman1978field}. How prevalent such processes are for other commonly studied species in cold dilute gases needs to be investigated. Additionally, our results demonstrate a convenient method of loading homonuclear molecular ions in an ion trap for further cold chemistry studies. A similar methodology can be explored in the future for loading heteronuclear molecular ions as well \cite{polak1981observation,gabbanini1991associative}.

\begin{acknowledgments}
We acknowledge support from the Ministry of Electronics and Information
Technology (MeitY), Government of India, under Centre
for Excellence in Quantum Technologies grant with Ref. No.
4(7)/2020-ITEA. The Workshop and Meena M S at RRI, and Jankee Upadhaya and Sudhir Kumar from RRCAT Indore for electronics support. EK acknowledges the DAE Raja Ramanna Fellowship.
\end{acknowledgments}

\section*{Supplemental material}

\section{Optical detection setup for the fluorescence detection of atoms.}
\label{sec:level1New}

\begin{figure}[h]

\centering
\hspace{-0.3 cm}
\includegraphics[scale=0.10]{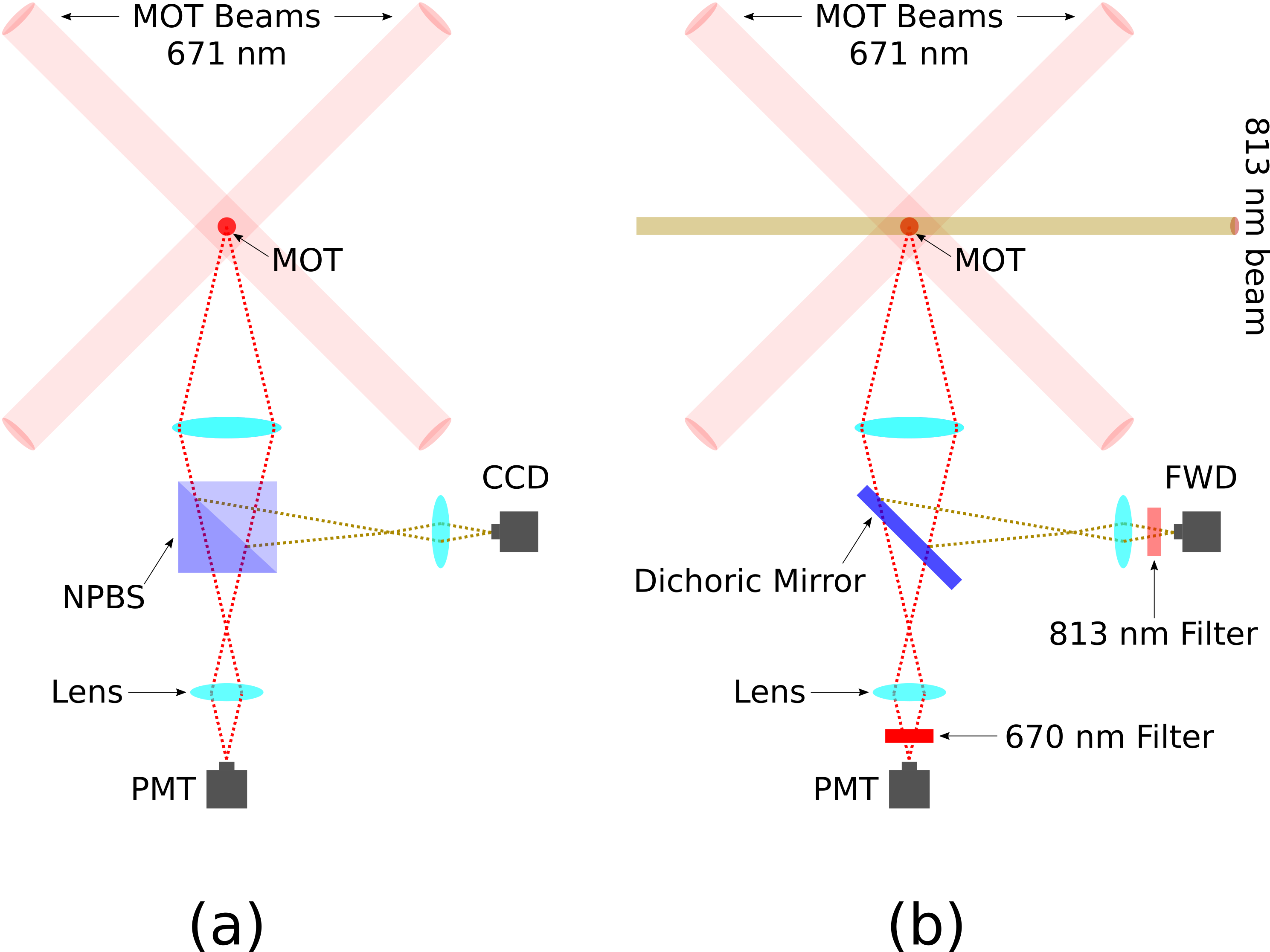}

\caption{(\textbf{a}) The characterization of the MOT (size, density, and temperature), as mentioned in the main text, was made using the $2P \rightarrow 2S$ fluorescence collected via a 2-inch lens, mounted outside the vacuum chamber. Using a non-polarizing beam splitter(NPBS) a part of the fluorescence is sent to a CCD camera and the other part to a photomultiplier tube(PMT). (\textbf{b}) The detection of the $2P \rightarrow 3S$ transition is made using the $3S \rightarrow 2P$ fluorescence. The NBPS is replaced by a dichroic mirror which separates the $3S$ emission line fluorescence from the MOT fluorescence and sends it to the femtowatt detector(FWD). The MOT fluorescence is detected by the PMT. The FWD and PMT additionally have 10 nm bandpass filters placed in front of them.\label{fig:SuppExpDiag}}
\end{figure}

\hspace{-0.35 cm}Fig \ref{fig:SuppExpDiag}(a) shows the optical detection setup for detecting the $2P \rightarrow 2S$ fluorescence for MOT characterization. Fig \ref{fig:SuppExpDiag}(b) shows the optical detection setup for detecting the $2P\rightarrow 3S$ excitation via the $3S \rightarrow 2P$ fluorescence.

\section{\label{sec:level2}$3S$ to $2P$ fluorescence} 

\begin{figure}[h!]
\centering
\hspace{-0.3 cm}
\includegraphics[scale=0.14]{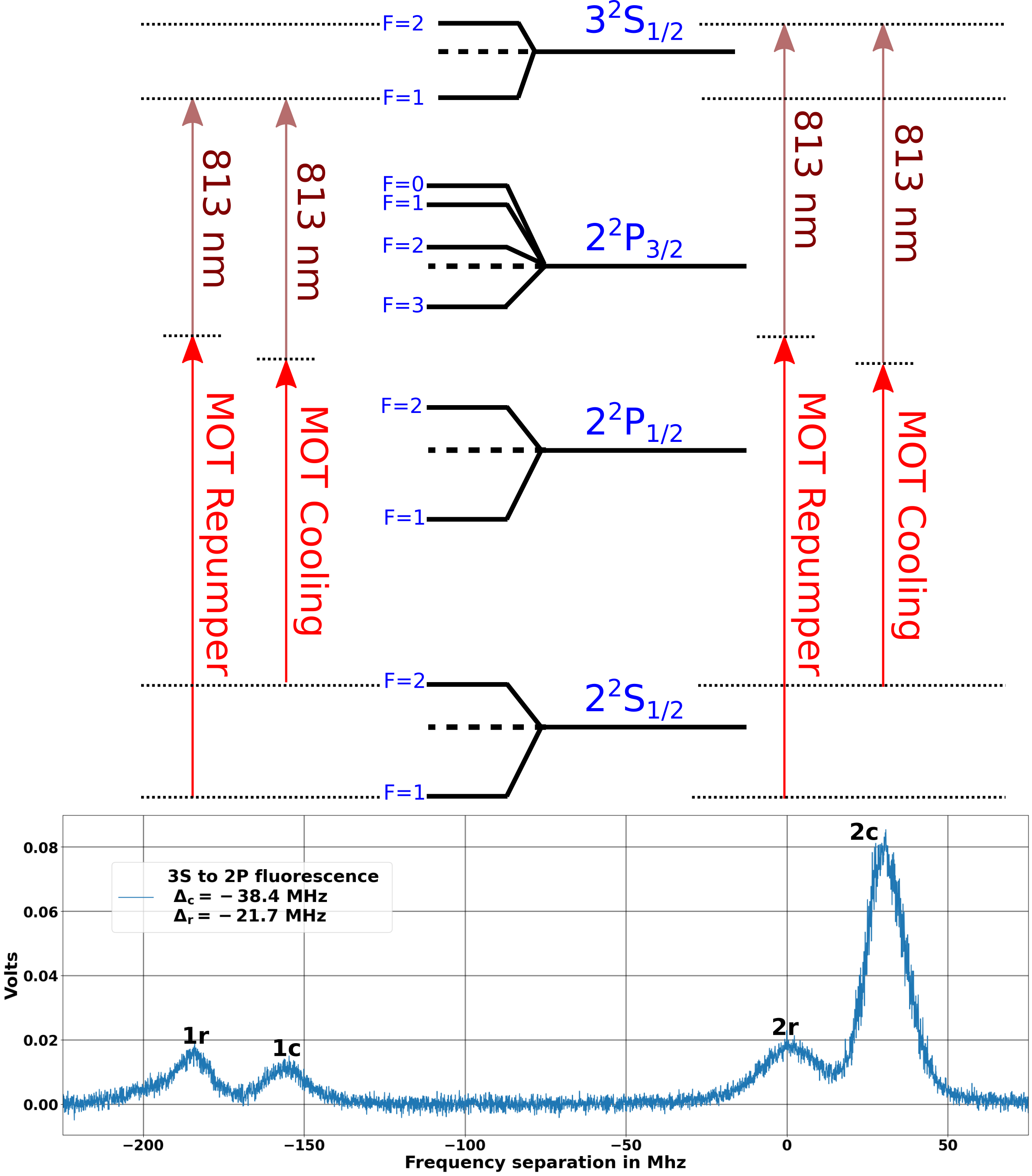}
\caption{$3S$ to $2P$ fluorescence peaks are shown aligned with their respective excitation by the MOT lasers and the 813 nm laser, as the 813 nm laser is scanned across the $2P_{3/2}\rightarrow 3S_{1/2}$ resonance. $\Delta_{c}$ and $\Delta_{r}$ are the detunings of the MOT cooling and repumper lasers from the $2S_{1/2}(F=2) \rightarrow 2P_{3/2}(F=3)$ and $2S_{1/2}(F=1) \rightarrow 2P_{3/2}(F=2)$ transitions.\label{fig:2p3sSpectrumSupplementary}}
\end{figure}

\hspace{-0.35 cm}Fig. \ref{fig:2p3sSpectrumSupplementary} shows a typical fluorescence spectrum obtained by scanning the 813~nm laser across the $2P_{3/2} \rightarrow 3S_{1/2}$ resonance. The fluorescence of atoms decaying from the $3S_{1/2}$ state to the intermediate $2P_{3/2}$ and $2P_{1/2}$ states is collected on a femtowatt detector (FWD) as shown in Fig. 1(b). The peaks labeled ''1c'' and ''2c'' arise from atoms decaying from the $3S_{1/2}(F=1)$ and $3S_{1/2}(F=2)$

\hspace{-0.35 cm}states respectively, when they were excited to these states by absorbing 2 photons, one from the MOT cooling laser and the other from the 813~nm laser. Likewise, the fluorescence peaks labeled ''1r'' and ''2r'' are due to atoms decaying from the $3S_{1/2}(F=1)$ and $3S_{1/2}(F=2)$ states respectively when they were excited to these states by absorbing 2 photons, one from the MOT repumper laser and the other from the 813~nm laser. The peak heights are different due to differences in the 2 photon transition strengths as different hyperfine states of the $2P_{3/2}$ manifold are involved in the transition. The 2c-1c and 2r-1r peak separations are in agreement with the $^7$Li hyperfine splitting of the $3S_{1/2}$ state \cite{kumar2017precise}. For all the measurements in the main text, the 813~nm laser was locked on the 2c peak. But we observe molecular ions at other peak values as well.

\section{\label{sec:level1}Detection of trapped ions} 

\hspace{-0.35 cm}For detecting trapped ions, the RF on the cylindrical electrodes is switched off at zero phase, and simultaneously a negative high voltage (HV) pulse is applied to the grid electrode to pull the ions towards the MCP (Fig. 1(a) in the main text). Pulsed HV is required 
to prevent the ions that have crossed the grid from being pulled back towards the grid. Therefore the duration ($t_{HV}$) and height ($V_{HV}$) of the pulse need to be adjusted accordingly. Additionally, the pulse duration and height can be adjusted so that ions of a certain charge to mass (Q/M) ratio reach the MCP but others do not, as the grid voltage pulls them back. We have used this for differentiating between the ions generated by the LED and the 813~nm laser in Section \ref{sec:levelDistinction} below and also in probing ion resonances using parametric excitation as described in Section \ref{sec:level3} below.

\section{\label{sec:levelDistinction}Configurations distinguishing the LED and 813~nm generated ions}

\hspace{-0.35 cm}Configuration A, in Table \ref{tab:table1} below, is the typical ion trapping and extraction configuration in our experimental setup. In this configuration, the ToF spectrum of the Li$^+$ ions generated by the UV LED and that of the 813~nm laser generated ions is respectively the green and black curves in Fig. 2 in the main text. Configurations B and C (Table \ref{tab:table1}) are designed to distinguish between the Li$^{+}$ ions, and the Li$_m^{+}$ ions which appear as a diffuse peak in the black curve in Fig. 2 in the main text.\\
In configuration B, the ion trap parameters are kept the same as in configuration A, but the height ($V_{HV}$) and duration ($t_{HV}$) of the negative HV pulse on the grid is increased so that the Li$^{+}$ ions do not reach the MCP. In configuration C, the negative high voltage pulse characteristics are kept the same as in configuration A, but the ion trap parameters are adjusted to lie outside the ion trap stability region
of Li$^{+}$. Configurations A and B are used in the parametric excitation measurements in Section \ref{sec:level3} below.

\begin{table}[h]
\caption{A, B, and C are configurations designed to distinguish between the LED and 813~nm generated ions. $V_{rf}$ and $\Omega_{rf}$ is respectively the rf voltage and angular frequency applied to one pair of diagonal rf electrodes and the other pair is grounded. $V_{ec}$ is the voltage applied to the endcaps. $V_{HV}$ and $t_{HV}$ is respectively the height and duration of the HV pulse on the grid electrode.}
\begin{ruledtabular}
\begin{tabular}{cccccc}
 &$V_{rf}$ &$\Omega_{rf}/2\pi$ &$V_{ec}$ &$V_{HV}$
 &$t_{HV}$\\
 &(V)&(kHz)&(V)&(V)
 &($\mu$s)\\
\hline
A& 70 & 1000 & 4 &-600
& 1.8\\
B& 70 & 1000 & 4 &-1000
& 2.4\\
C& 76 & 500 & 10 &-600
& 1.8\\
\end{tabular}
\end{ruledtabular}
\label{tab:table1}
\end{table}

\section{\label{sec:level3}Ion resonances using parametric excitation.}

\begin{figure}[t]
\centering
\includegraphics[scale=0.50]{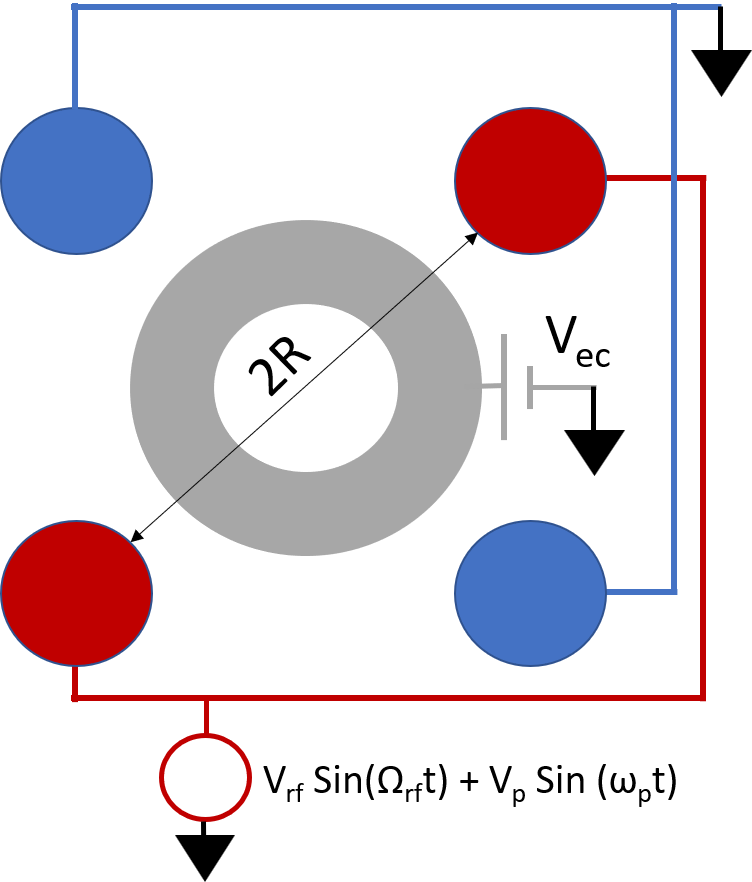}
\caption{Axial view of the linear Paul trap. The annular disc at the center is the endcap on which a DC voltage $V_{ec}$ is applied. RF signal of amplitude $V_{rf}$ and angular frequency $\Omega_{rf}$ plus the parametric drive of amplitude $V_{p}$ and angular frequency $\omega_{p}$ is applied on one diagonal pair of electrodes while the other pair is grounded.  \label{fig:TrapAxialView}}
\end{figure}

\begin{figure}[h!]
\centering
\includegraphics[scale=0.18]{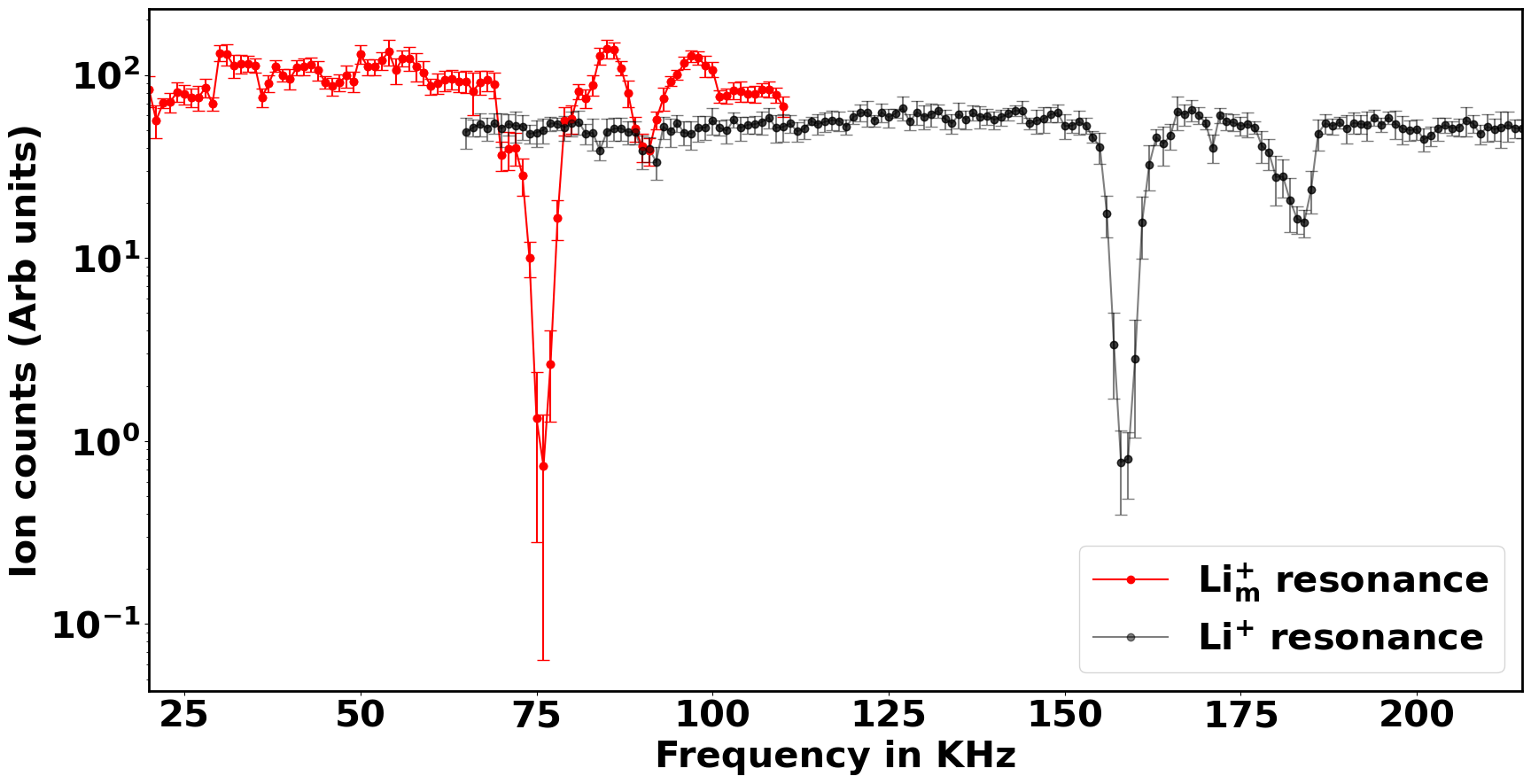}
\caption{Measurement of the resonances of Li$^+$ and Li$_{m}^{+}$ at $V_{rf}$ = 70~V and $\Omega_{rf}/2\pi = 1000$~kHz and $V_{ec} = 1$~V. The peak to peak parametric drive amplitude, in units of milli volts ($mV_{pp}$), is 180 $mV_{pp}$ and 67.5 $mV_{pp}$ respectively, for the Li$^{+}$ and Li$_{m}^{+}$. The amplitude of the parametric drive is kept as low as possible so that only strong resonances are visible. The frequency of the parametric drive is varied in steps of 1~kHz.\label{fig:Parametric}}
\end{figure}

\hspace{-0.35 cm}Since the ToF's is not enough to determine the mass of Li$_{m}^{+}$, we use the method of parametric excitation (PE) \cite{zhao2002parametric} to determine the resonant frequency of Li$_{m}^{+}$ and compare it to that of Li$^+$. A small amplitude ($V_{p}$) parametric drive of a particular frequency ($\omega_{p} = 2\pi f_p$) is mixed with $V_{rf}$ as shown in Fig. \ref{fig:TrapAxialView}. The ions, generated either via the LED or the 813~nm laser, are loaded in the ion trap for 0.5~s. Beyond this time they are held in the ion trap for another 1.5~s before getting extracted toward the MCP. This measurement is done at different parametric drive frequencies and in each case, the number of ions reaching the MCP is recorded. For comparing the resonant frequencies of the two ions, the ion trap parameters need to be the same. Configurations A and B (see Table \ref{tab:table1}) have the same ion trap parameters. For the LED generated Li$^+$ ions, we measure the resonant frequencies in configuration A, where the HV pulse characteristics are well suited for the detection of Li$^+$. However, for the Li$_{m}^{+}$, generated by the 813~nm laser, we performed measurements in configuration B where the HV pulse is adjusted so that the Li$^+$ ions do not reach the MCP. This ensures that Li$^+$ ions, generated via the 813~nm, do not appear as a source of background noise in the measurement of the resonant frequency of Li$_{m}^{+}$. For reasons explained below, we have performed measurements at $V_{rf} = 70$~V and $\Omega_{rf}/2\pi= 1000$~kHz but at three different values of $V_{ec}$. Table $\textrm{I}$, in the main text, compares the most prominent resonance for Li$^+$ with that of Li$_{m}^{+}$ at different endcap voltages. Fig. \ref{fig:Parametric} shows the measurement achieved at $V_{ec} = 1$~V.\\
We recall that in the adiabatic approximation \cite{major2005charged,drakoudis2006instabilities,sinhal2023molecular}, the ion secular frequencies in an ideal linear Paul trap, with no DC offset on the RF electrodes, are given by 
\begin{equation}
    \omega_{u} \approx  \frac{\Omega_{rf}}{2}\sqrt{\frac{q^{2}_{u}}{2} + a_{u}},
\end{equation}

\hspace{-0.35 cm}with  $u \equiv x,y,z$, and $(x,y)$ and $z$ being the radial and axial directions respectively, and 
\begin{eqnarray}
    q_{x} = -q_{y} = \frac{2 Q V_{rf}}{M r_{0}^{2} \Omega_{rf}^{2}}, \\
    q_{z} = 0,\\
    a_{x} = a_{y} = -\frac{a_{z}}{2} = -\frac{4 \kappa Q V_{ec}}{M z_{0}^{2} \Omega_{rf}^{2}}.
\end{eqnarray}
Here $2r_0$ represents the surface-to-surface separation between diagonal RF electrodes(see Fig. \ref{fig:TrapAxialView}), $2z_0$ is a similar separation between the endcaps, and $\kappa$ is a geometrical factor. $Q$ and $M$ are the charge and mass of the ion species. In the limit of $V_{ec} \rightarrow 0$, we have $a_{u} \rightarrow 0$, $\omega_{u} \rightarrow  q_{u}\frac{\Omega_{rf}}{2\sqrt{2}}$. The ratio of the radial secular frequency ($\omega_{x/y}$) of two different ions is approximately equal to their ($Q/M$) ratios. This holds not only for the radial secular frequencies but also for any resonance which is not purely axial (\textit{i.e.} which are multiples of only $\omega_{z}$) and are of the type $\omega_{ion} = n_{1} \times \omega_{x/y} \pm n_{2} \times \omega_{z}$, where $n_{1}$ and $n_{2}$ are numbers and $n_{1} \neq 0$. This is because $\omega_{z} \rightarrow 0$ as $V_{ec} \rightarrow 0$. The linear Paul trap essentially behaves like a quadrupole mass spectrometer \cite{collings2003resonant} in the limit of $V_{ec}\rightarrow 0$.\\
The resonances in Table \ref{tab:fpe}, in the main text, are radial resonances. The ratio of these resonant frequencies converges to 2.0 as the endcap voltage is lowered. This suggests that $(Q/M)_{\textrm{Li}_{m}^{+}} = \frac{1}{2} (Q/M)_{\textrm{Li}^{+}}$ and that the ion generated by the 813~nm laser, which appears as a broad diffuse peak in the time of flight spectrum, is Li$_{2}^{+}$.

\section{\label{sec:level4}Counting atomic and molecular ions in the histograms in Fig. 4 in the main text}

\begin{figure}[h!]
\centering
\includegraphics[scale=0.57]{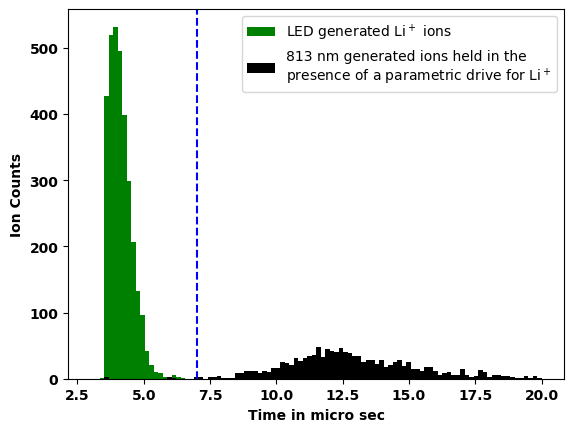}
\caption{ToF distribution of the Li$^+$ and Li$_2^+$ ions, in configuration A. The blue dashed vertical line, at 7~$\mu$s, is the cutoff ToF. Ions before and beyond this ToF are counted as Li$^+$ and Li$_2^+$ respectively. \label{fig:cutOffTime}}
\end{figure}

\hspace{-0.35 cm}In Fig. 2, main text, since the diffuse ToF structure of Li$_2^+$ has slight overlap with the ToF distribution of Li$^+$ (green curve), we adopted the following method to count the number of Li$^+$ and Li$_{2}^+$ ions in the histograms in Fig. 4(a) in the main text. We first build the ToF distribution of just the Li$^+$ ions. For this purpose, for 10~ms, we load Li$^+$ ions in the ion trap by ionizing MOT atoms in the $2P_{3/2}$ state via the UV LED. The ions are held in the ion trap, operated in configuration A (see Table \ref{tab:table1}), for another 500~ms, and are then extracted onto the MCP. 500 of such measurements are performed and the ToF distribution is recorded for each measurement. The distribution, over all these 500 measurements, is shown in Fig. \ref{fig:cutOffTime}.\\
For extracting the actual ToF distribution of just the molecular ions, we need to remove Li$^+$ ions which are generated along the way and are trapped together with Li$_2^+$. For that purpose, we again use PE to remove Li$^+$. For each measurement, the MOT is loaded for 20~s. Beyond this time, the MOT light is switched off. But just before that, an amount of 5~mW$/$cm$^2$ of the 813~nm light is incident on the MOT for 10 ms to create ions. The ions are held in the ion trap, for 2~s, in the presence of a parametric drive for Li$^+$, and are then extracted onto the MCP. These measurements are made in configuration A (see Table \ref{tab:table1}). Fig. \ref{fig:cutOffTime} shows the ToF distribution over 500 of such measurements.\\
We see that the ToF distributions of Li$^+$ and Li$_2^+$ are well separated with a very small overlap. We choose 7~$\mu$s as the cutoff time (blue dashed vertical line in Fig. \ref{fig:cutOffTime}) for counting the number of Li$^+$ and Li$_2^+$ ions in Fig. 4(a) in the main text. All ions with ToF less than 7~$\mu$s are counted as Li$^+$ and the ones beyond this time are counted as Li$_2^+$.

\section{Associative ionization rate}

\begin{figure}[h!]
\centering
\includegraphics[scale=0.32]{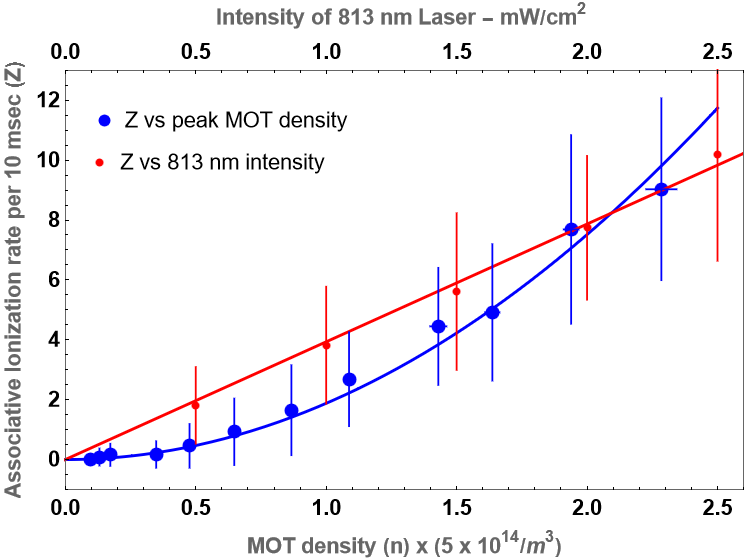}
\caption{Variation of the associative ionization rate with the intensity of the 813~nm laser and with the peak MOT density. We note that the data just shows the trend of the variation of the associative ionization rate and not absolute numbers: they can be different in reality depending on the efficiency of detection, and on the overlap between the MOT (with FWHM of $\approx 1$~mm) and the 813~nm beam (with diameter of $\approx 1$~ mm) for a particular measurement set.\label{fig:IonizationRateWithDensityAndIntensity}}
\end{figure}

\hspace{-0.35 cm}For measuring the rate of associative ionization (AI), the 813~nm laser is continuously illuminating the $^7$Li MOT atoms. Further, the ion trap electrodes are configured to continuously extract the produced ions onto the MCP. For determining the rate at which the ions are hitting the MCP the oscilloscope continuously records a total of 100 frames, each of 10~ms duration containing electronic pulses corresponding to ions hitting the MCP. For recording the variation of the AI rate with the MOT density, the intensity of the 813~nm laser is kept to $\approx 1$~mW/cm$^2$. For measuring the variation of the rate with intensity the MOT is maintained at a density of $\approx 10^{15}$ atoms/m$^3$. The red data with error bars in Fig. \ref{fig:IonizationRateWithDensityAndIntensity} shows the variation of the AI rate with the 813~nm intensity and is fitted to a linear function. The blue data with error bars in Fig. \ref{fig:IonizationRateWithDensityAndIntensity} shows the variation of the AI rate with the MOT density and is fitted to a quadratic function.

\bibliography{MolecularIonsMain}

\end{document}